\documentclass[aps,prl,twocolumn,showpacs,amssymb,amsmath]{revtex4}
\usepackage{graphicx}
\usepackage{epsfig}
\usepackage{bm}

\begin{document}

\title{Domain regime in two-dimensional disordered vortex matter} 
\author{Mahesh Chandran} 
\altaffiliation{Present address: Materials Research Laboratory, John F Welch Technology 
Centre, Bangalore, 560066.}
\email{mahesh.chandran@ge.com} 
\author{R. T. Scalettar}
\author{G. T. Zim\'{a}nyi}
\affiliation{Department of Physics, University of California, Davis, California 95616.}

\date{\today}

\begin{abstract} 

A detailed numerical study of the real space configuration of vortices in
disordered superconductors using $2D$ London-Langevin model is presented. The
magnetic field $B$ is varied between 0 and $B_{c2}$ for various pinning strengths
$\Delta$. For weak pinning, an inhomogeneous disordered vortex matter is observed,
in which the topologically ordered vortex lattice survives in large domains. The
majority of the dislocations in this state are confined to the grain
boundaries/domain walls. Such quasi-ordered configurations are observed in the
intermediate fields, and we refer it as the domain regime (DR). The DR is distinct
from the low-field and the high-fields amorphous regimes which are characterized by
a homogeneous distribution of defects over the entire system. Analysis of the real
space configuration suggests domain wall roughening as a possible mechanism for the
crossover from the DR to the high-field amorphous regime. The DR also shows a sharp
crossover to the high temperature vortex liquid phase. The domain size distribution
and the roughness exponent of the lattice in the DR are also calculated. The
results are compared with some of the recent Bitter decoration experiments.

\end{abstract}


\maketitle

\section{Introduction}

The vortex state in type-II superconductors is a paradigm for studying the
effect of quenched disorder in condensed matter. Over the last decade, much of
the effort has been spent on characterizing the various phases of the vortex
state as a function of the magnetic field $B$ and the temperature $T$. For $3D$
vortex system\cite{comment0}, three phases have been identified
unambiguously~\cite{blatter,gia1,nattermann1}:  the Bragg glass (BG) with
quasi-long range order, the amorphous vortex glass (VG), and the vortex liquid
(VL). The VG and the VL are distinguished by their superconducting and ohmic
responses, respectively. Experiments in high-$T_c$ superconductors suggest that
the BG phase appears in the low-$B$ and low-$T$ region whereas the VG phase
occupies the high-$B$ and low-$T$ region of the $B$-$T$ phase diagram. The VL
phase appears close to the upper critical field $B_{c2}(T)$.

The first detailed calculation of the real space structure of the vortex lattice
in the presence of quenched impurities\cite{comment1} was carried out by Larkin and
Ovchinnikov\cite{larkin} (LO). The LO theory assumes that for the weak pinning,
the vortex lattice is coherently pinned within a volume $V_{c}$. Beyond $V_{c}$,
the effect of impurities dominates and the long range positional order is lost.
It was further proposed that when the vortex displacement becomes of the order of
lattice constant, topological defects (dislocations in $D=2$ and dislocation
loops in $D=3$) are generated\cite{fisher}. Later calculations by Giamarchi and
Le Doussal~\cite{gia1,nattermann3} (GLD) showed that the LO calculation
overestimates the effect of impurities at large distances and the vortex
displacement grows only logarithmically (for $3D$ vortex system). The positional
correlation decays as $C(r)\sim\frac{1}{r^{\eta}}$, where $\eta$ is a
non-universal exponent\cite{emig}. The quasi-long range $C(r)$ leads to a
topologically ordered phase (the Bragg glass), which is stable with respect to
the formation of defects\cite{fisher1}. On increasing $B$ (or the pinning
strength), the LO and GLD theories predict proliferation of topological defects,
thus forming the VG at high fields\cite{comment2}. The transition between the BG
and the high field VG is predicted to be of
first-order\cite{gia3,ertas,vinokur,koshelev}. The BG also undergoes a melting
transition to VL on increasing $T$.

In $D=2$, the BG phase is unstable to the formation of dislocations and the
positional quasi-long range order is destroyed\cite{zeng,gingras}. However, for
weak pinning and at low temperatures, the unbound dislocations appear only at
large length scale $\xi_{D}\gg R_a$, where $R_a$ is the ``Random Manifold''
length scale and is the distance at which positional correlation begins to
decay\cite{doussal}. On length scales shorter than $\xi_D$, the topologically
ordered lattice forms a quasi-Bragg glass (qBG). Such a qBG shows an
exponentially sharp crossover to the high temperature VL phase\cite{carpentier1},
reminiscent of the ``melting'' transition of the pure system. A similar
exponential crossover was proposed between the qBG and the VG phase as a function
of $B$, or pinning strength.

The real space structure of the vortex system has been studied used neutron
diffraction\cite{journard,klein,ling} and Bitter decoration of
vortices\cite{grier,march,menghini,fasano}. The latter technique allows direct
visualization of the large scale structure of the configuration and hence enables
one to analyze the role of topological defects on the decay of translational
order. Recent decoration experiments~\cite{menghini,fasano} of NbSe$_2$ have
raised some important issues concerning the nature of the disordered phase.
Previous transport measurements\cite{paltiel} on the same samples of NbSe$_2$
suggested an order-disorder transition on increasing $B$ (or $T$). Fasano {\it et
al.} showed that the spatial configuration of vortices does not show any
significant difference between the ordered and the disordered vortex phases
identified in Ref.\cite{paltiel}. More importantly, both phases were found to be
polycrystalline with dislocations forming grain boundaries. Within each grain,
the lattice shows significant bond orientational order. This is in contrast to
the naive theoretical picture of the ordered phase which expects a
dislocation-free configuration, and the disordered phase in which the
distribution of the dislocations is expected to be homogeneous.

In this paper, we analyze in detail the real space configuration of the
disordered phase using numerical simulation of a $2D$ vortex system at $T=0$.
The magnetic field $B$ is varied over a wide range for various values of the
pinning strength $\Delta$. The real space configuration shows that for the
intermediate field range, the system shows inhomogeneous disordering. The
majority of the dislocations are confined to the grain boundaries which forms
the domain wall between regions of ordered lattice. The domain size and its
distribution is dependent on $B$ and $\Delta$. We refer the intermediate fields
in which the vortex state is quasi-ordered as the domain regime (DR). The DR is
distinct from the amorphous regime at low-fields and high-fields, where the
defects appear at a length scale $\sim a_{0}$ (lattice constant) and its
distribution is homogeneous over the entire system. Analysis of the real space 
images suggests domain wall roughening as a possible mechanism for the
crossover between the DR and the high-fields amorphous regime. We also obtained
the roughening exponent $\zeta$ of the vortex lattice in the domain regime.
Finite temperature simulation shows that the domain regime undergoes a sharp
crossover to the high-$T$ liquid phase, which is reminiscent of the thermal
melting in the pure vortex system.

The paper is organized as follows: in section II, we discuss the simulation
approach in detail. The results and the analysis of the real space configuration
are presented in section III, followed by conclusions in section IV.

\section{Simulation method}

We consider a $2D$ cross-section perpendicular to the magnetic field ${\bf
B}=B\hat{\bf z}$ of a bulk type-II superconductor in the mixed state. Within
London's approximation, the vortex can be considered as a point particle with
the dynamics governed by an overdamped equation of motion
\begin{eqnarray} 
\eta\frac{d{\bf r}_{i}}{dt} = -\sum_{j\neq i} \nabla U^{v}({\bf r}_{i}-{\bf r}_{j}) - \sum_{k} 
\nabla U^{p}({\bf r}_{i}-{\bf R}_{k}) \nonumber \\
+ {\bf F}_{ext} + {\bm \zeta}_{i}(t).
\end{eqnarray} 
Here, $\eta$ is the flux-flow viscosity. On the left hand side, the first term
represents the inter-vortex interaction
$U^{v}(r)=\frac{\phi_{0}^{2}}{8\pi^{2}\lambda^{2}} K_{0}(\tilde{r}/\lambda)$,
where $K_0$ is the zeroth-order Bessel function, and $\tilde{r} =
(r^{2}+2\xi^{2})^{1/2}$. $\phi_{0}$ is the flux quantum, and the $\lambda$ and
$\xi$ are the penetration depth and the coherence length of the superconductor,
respectively. This form of the inter-vortex interaction includes the finite core
size of the vortex~\cite{clem}. The second term represents vortex pinning by
parabolic potential wells, where $U^{p}(r)=U_{0}(\frac{r^{2}}{r_{p}^{2}}-1)$ for
$r < r_{p}$, and 0 otherwise. The pinning centers are randomly located at
positions ${\bf R}_{k}$ in the simulation box. The third term ${\bf
F}_{ext}=\frac{1}{c}{\bf J}\times \phi_{0}\hat{\bf z}$ is the Lorentz force
experienced by the vortex due to the transport current density ${\bf J}$. The
thermal noise is represented by ${\bm \zeta}$ with
$\langle\zeta_{i,p}(t)\rangle\!=\!0$, and
$\langle\zeta_{i,p}(t)\zeta_{j,p'}(t')\rangle\!=\!
2k_{B}T\eta\delta_{ij}\delta_{pp'}\delta(t-t')$, where $T$ is the temperature,
$k_{B}$ is the Boltzmann constant, and $p,p'\!=\!x,\!y$. The length is in units
of $\lambda(B=0, T=0)\!=\!\lambda_{0}$, and the temperature $T$ is in units of
$\lambda_{0}f_{0}/k_{B}$, where
$f_{0}\!=\!\!\frac{\phi_{0}^{2}}{8\pi^{2}\lambda_{0}^{3}}$. The current density
$J$ and the velocity $v$ of the vortices are in units of $cf_{0}/\phi_{0}$ and
$f_{0}/\eta$, respectively. Also, the prefactor for the pinning potential 
$U_{0}$ is scaled by $f_{0}/\lambda_{0}$.

We use the reduced magnetic field $b=B/B_{c2}$, where the upper critical field
$B_{c2}\!=\!\frac{\phi_{0}}{2\pi\xi_{0}^{2}}$ and
$\xi_{0}\!=\!\xi(B\!=\!0,T\!=\!0)$. The $b$ is calculated from the lattice
constant
$\frac{a_{0}}{\lambda_0}\!=\!(\frac{4\pi}{\sqrt{3}})^{\frac{1}{2}}(\frac{1}{\kappa^{2}b})^{\frac{1}{2}}$.
The Ginzburg-Landau parameter $\kappa\!=\!\frac{\lambda}{\xi}$ is an input to
the simulation. The magnetic field dependence of the length scales $\lambda$ and
$\xi$ follows the relation $\lambda(b)=f(b)\lambda_{0}$ and
$\xi(b)=f(b)\xi_{0}$, respectively. The renormalization factor
$f(b)=(1-b^{2})^{-\frac{1}{2}}$. This form of the renormalization factor is
similar to the temperature dependence of $\xi$ and $\lambda$ in the
Ginzburg-Landau theory\cite{tinkham} with $T/T_{c}$ replaced by
$(B/B_{c2})^{2}$. Similar form of the renormalization factor for $\lambda$ have
been used in Ref.\cite{ryu}. The parameters used in the simulation are
$\kappa=10$ and $\lambda_{0}=1000\AA$, which are close to the values for the
low-$T_{c}$ superconductors, particularly NbSe$_2$. Periodic boundary conditions
are imposed in both directions, and the minimum image convention is followed.  
The magnetic field $b$ is varied by changing the size of the simulation box,
keeping the number of vortices $N_{v}=4096$ fixed. Simulations were also
performed using $N_{v}$ between 800-1200, and for some parameters $N_{v}=6400$
was used to check for the finite size effects. The $U_{0}$ is distributed
randomly between $\Delta\pm 0.01$, where $\Delta = \langle U_{0}\rangle$. The
range of the pinning potential $r_{p}=\xi_{0}$. In this paper, we present
results for pin density $n_{p}=2.315/\lambda_{0}^{2}$. For $T=0$, Eq.(1) is time
integrated by the predictor-corrector scheme, and the finite temperature
simulation is performed using Heun's method\cite{greiner}. The simulation at
high vortex densities requires long computational time and parallel algorithms
were employed to reduce the run time. Details of the implementation of the
parallel algorithms can be found in Ref.\cite{chandran1}.

The real space configuration is characterized by the topological defect density
$n_{d}/\lambda_{0}^{2}$ (number of defects per unit area of the simulation box).
Below, we also use the defect fraction $f_{d}$, which is defined as the number
of defects per vortex. The defects are defined as vortices with coordination
number other than 6 and are identified by Delaunay triangulation of the real
space position of the vortices. In $2D$ systems, the vortices with coordination
number 5 and 7 are disclinations. A 5-disclination and a 7-disclination
separated by a distance $a_{0}$ forms a bound pair which is an edge dislocation.
Over most of the field range, the fraction of free disclinations is negligibly
small and the majority of the defects are edge dislocations. Hence, the defect
density $n_{d}$ is approximately twice the dislocation density in the system.
The hexatic order in the system is quantified by the six-fold orientational
order parameter $\Psi_{6} = |\sum_{\langle ij\rangle}e^{6\theta_{ij}}|$, where
$\theta_{ij}$ is the angle between the nearest neighbor vortices relative to a
reference axis.

The simulation is performed by two different methods. In the first method, we
start with a perfect vortex lattice and the driving current $I(\propto J)$ is
reduced to 0 from a value much greater than the depinning current $I_{c}$. This
is referred as the current annealing (CA) method. In the second method, the
conventional thermal annealing (TA) is applied wherein the temperature $T$ is
reduced to 0 in small steps from the high temperature liquid phase (also known as
simulated annealing). Experimentally, the TA is equivalent to the field cooling
procedure. We have shown previously\cite{chandran2} that the configuration
obtained by CA is stable to small perturbations compared to the configuration
obtained by TA. This is also supported by experiments\cite{henderson}, which show
that the field cooled state is unstable to small driving force $I\ll I_{c}$ and a
stable configuration is obtained when the system is brought to rest after driven
with $I\gg I_{c}$. The two methods, CA and TA, are compared in section III. B.

\begin{figure}[hbt] 
\epsfxsize=3.2in
\epsfysize=2.3in 
\includegraphics[width=230pt]{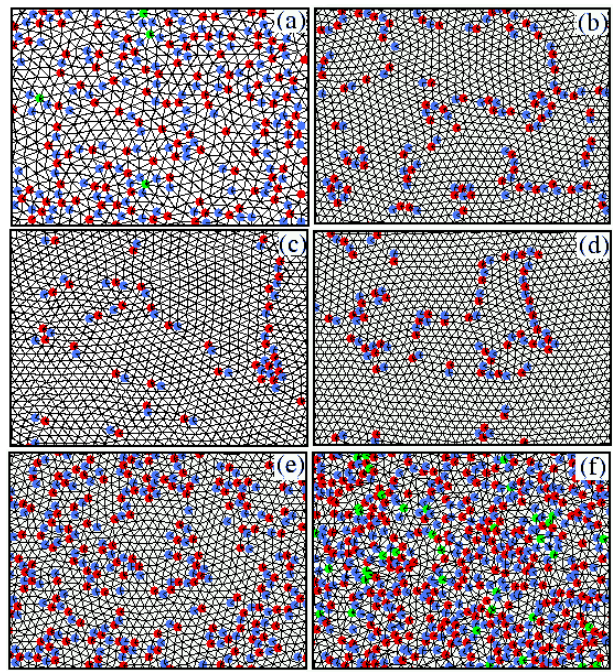} 
\caption{(Color Online) The real space configuration of vortices in a region
of the simulation box. The values of $b$ are (a) 0.1, (b) 0.4, (c) 0.5, (d)
0.6, (e) 0.8, and (f) 0.9. The black (red) and gray (blue) dots denote 
vortices with 7 and 5 neighbors, respectively\cite{comment-color}. For $b=0.1$, 
$N_{v}=900$, and for the rest of the images $N_{v}=4096$. The pinning strength 
$\Delta=0.02$.}
\end{figure}

\section{Results and discussions}

\subsection{Zero temperature simulation}

In this section, we analyze the zero-temperature configurations obtained by the
current annealing method. The system is slowly brought to rest across the
depinning current for each value of the magnetic field $b$. In the absence of
thermal fluctuations, the vortex configuration is determined by the balance
between the long range elastic force and the pinning force. We first show the
real space images of the configuration as the magnetic field $b$ is increased.

\subsubsection{Real space configuration}

Fig.1 shows the Delaunay triangulation in a region of the simulation box for
various values of the magnetic field. The pinning strength $\Delta=0.02$, and
$N_{v}=4096$, except for $b=0.1$ for which $N_{v}=900$. At small fields
$b\lesssim 0.1$, the defect distribution is homogeneous over the entire system
and the configuration is amorphous. The defect fraction (number of defects per
vortex) $f_{d}>0.35$ at these low fields.

With increasing $b$, small regions of ordered lattice start appearing. This can be
seen for $b=0.1$ in which ordered lattice is formed in regions less than 3-4$a_{0}$
wide. For $b\gtrsim 0.2$, the defect distribution becomes inhomogeneous. The
dislocations come closer to form a network of grain boundaries across the system.
For $b=0.4$ and $b=0.5$, we find that $\approx 90\%$ of the dislocations in the
system are confined to the grain boundaries whereas $\approx 10\%$ of the
dislocations are free within the domains. We also find that $\approx 5\%$ of the
dislocations {\em within} the grain boundaries unbind into disclinations, which
occurs generally at the intersection of the grain boundaries. Though the free
disclinations are absent in the system, it does not lead to a long range hexatic
order in the system. For $b=0.6$, we find $\Psi_{6}\approx 0.14$, and for other
values of the field $\Psi_{6} < 0.05$ (for a perfect vortex lattice,
$\Psi_{6}=1$). The small value of $\Psi_{6}$ is caused by the random
orientation of the domains which destroys the long range orientational order.

We call the intermediate field range in which the system breaks into regions of
ordered lattice the domain regime (DR). The domain regime is configurationally
distinct from the conventional picture of a disordered state for which the
distribution of topological defects is homogeneous. In the DR, the system is
quasi-ordered on the length scale of the domain size $R_{d}$. The vortex lattice
shows translational and orientational order {\em within} the domains, even though
the long range order is absent in the system. Fig.1(b)-(d) shows real space
configurations in the DR as $b$ is increased.

\begin{figure}[bt]
\epsfxsize=3.2in 
\epsfysize=2.3in 
\includegraphics[width=230pt]{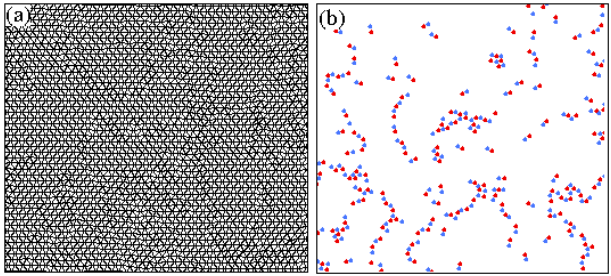} 
\caption{(Color Online) The real space configuration of vortices in a region 
of the simulation box for $\Delta=$0.01 (a), and 0.02 (b). The $N_{v}=6400$. 
In (b), only the defects are shown.} 
\end{figure}

For $b>0.6$, the defect density increases rapidly with the concomitant decrease
in the domain size. Small domains of ordered lattice of width 3-4$a_0$ can be
seen until $b\approx 0.8$, as evident from Fig.1(e). Increasing $b\gtrsim 0.8$,
the system becomes amorphous with an average distance $\sim a_{0}$ between the
defects. The defect distribution is homogeneous throughout the system, and the
configuration is similar to a frozen liquid. A typical real space configuration
is shown in Fig.1(f). The fraction of free disclinations is significantly higher
than that observed in the DR but it is difficult to isolate them from the dense
network of defects. The vortices with coordination number 4 and 8 accounts for
$\approx $6-8\% of the total defects. Overall, the real space images in Fig.1
suggests a reentrant change in the configuration, from a low-field amorphous to
an intermediate field quasi-ordered state, which finally crosses over to a
high-field amorphous state.

In the DR, the average domain size $R_{d}$ is dependent on $b$ and $\Delta$.
$R_{d}\approx 5-7a_{0}$ for low fields and increases in the intermediate field
range. For $b=0.6$ and $\Delta=0.02$, the size of some domains exceeds $20a_{0}$.
By decreasing the pinning strength $\Delta$ to 0.01, we find a remarkably well
ordered lattice with no topological defects for the system size $N_{v}=6400$, as
shown in Fig.2(a). This suggests that for sufficiently weak pinning strength,
large domains of ordered lattice, comparable to sample size in typical
experiments, can exist in $2D$. Fig.2(a) should be compared with Fig.2(b) which
shows the meandering grain boundaries formed by the defects for $\Delta=0.02$.
With increasing $\Delta$, $R_{d}$ decreases from $\approx 20a_{0}$ for
$\Delta=0.02$ to $\approx 3$-$5a_{0}$ for $\Delta=0.075$. This is shown in Fig.3.
For strong pinning, the dislocations tend to cluster in some regions implying
that individual pinning centers locally tear the vortex lattice. With increasing
$\Delta$, the field range over which the DR exists is reduced and in the strong
pinning limit, the system is amorphous for all values of $b$.

\begin{figure}[bt]
\epsfxsize=3.2in 
\epsfysize=2.3in 
\includegraphics[width=230pt]{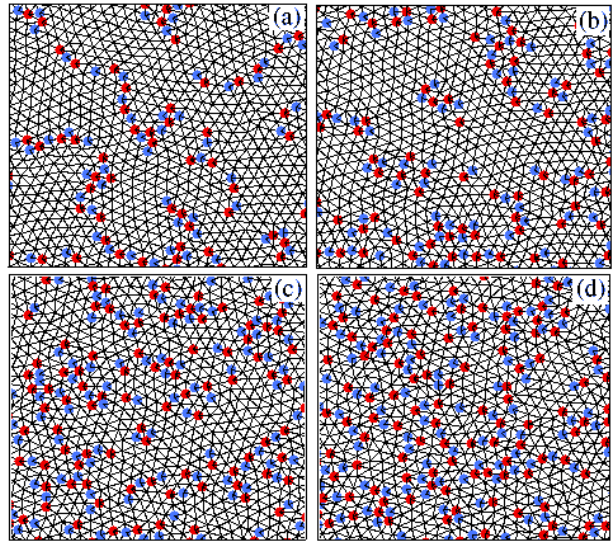} 
\caption{(Color Online) The vortex configuration in a region of the simulation 
box for $b=0.6$ and for $\Delta=0.03$ (a), 0.04 (b), 0.05 (c), and 0.075 (d). 
The number of vortices $N_{v}=4096$.} 
\end{figure}

In recent experiments, the changes in the real space configuration of vortices
were studied in weak pinning NbSe$_2$ samples across the order-disorder
transition by the Bitter decoration technique\cite{menghini,fasano}. The
order-disorder transition was previously identified in transport measurements and
have been speculated to underlie the peak effect in the critical current
density\cite{paltiel}. The decoration images show that the vortices form large
ordered domains. The domains are separated by domain walls, which are defined by
chains of dislocations. This domain formation is present throughout the B-T plane
(below the melting line), hence the authors summarized their findings as the
``absence of amorphous vortex matter''. Fasano {\it et al.} found that
$\approx$85-90\% of the defects are in the grain boundaries, whereas the
remaining defects are isolated dislocations. All of these findings are consistent
with our numerical findings and estimates in the intermediate field range for
$\Delta=0.02$.

\subsubsection{Domain size distribution}

A useful quantity to characterize the DR is the distribution of the domain size
$N(s_{d})$, where the area of the domain $s_{d}$ is in units of $a_{0}^{2}$.
Unlike in lattice models, extracting $N(s_{d})$ in models with continuous
symmetry is not straightforward. The lattice vectors can change continuously from
domain to neighboring domain {\em without} nucleating defects, which makes it
difficult to define the domain wall. In many cases, the domain walls, which are
formed by the grain boundaries are not closed. Analysis of the real space
configuration suggests that the domain walls are generally composed of two types
of grain boundaries, depending on the misorientation angle $\theta_d$ between the
neighboring domains. For the small angle grain boundaries, $\theta_{d}\sim
10^{\circ}$-$16^{\circ}$, and the dislocations are separated by 3-5$a_0$. In
large angle grain boundaries, the dislocations form closely packed string-like
structures, and $\theta_{d}>20^{\circ}$. Typical domains and domain walls formed
by the grain boundaries are shown in Fig.4.

\begin{figure}[bt]
\epsfxsize=3.2in 
\epsfysize=2.3in 
\includegraphics[width=230pt]{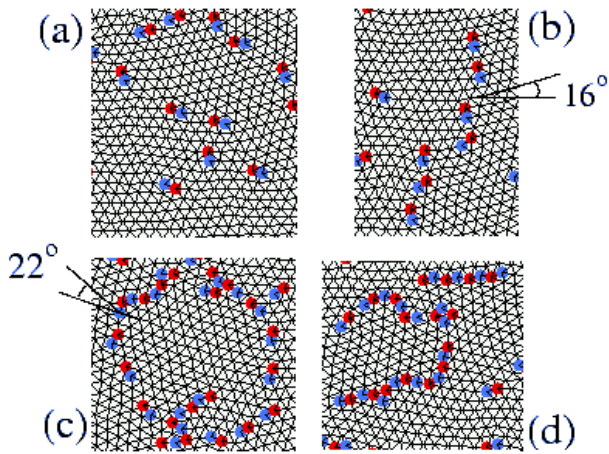} 
\caption{(Color Online) The domain walls in the real space configuration. (a) 
and (b) shows small angle grain boundaries, whereas large angle grain 
boundaries can be seen in (c) and (d). The magnetic field $b=0.50$ for (a) and 
(d), and 0.65 for (b) and (c).} 
\end{figure}

To extract $N(s_{d})$, we used the following procedure to define domain walls in
regions where the dislocations are apart. The mis-orientation angle $\theta_{d}$
is obtained between successive vortices along one of the lattice vectors
connecting neighboring domains. If $\theta_{d}$ is between $12^{\circ}$ and
$18^{\circ}$, that vortex is considered as part of the domain wall. With this
method, the domain boundary in many cases could be determined with reasonable
accuracy. This procedure is intended to be instructive rather than decisive, as
it contains some arbitrariness. For example, in some regions the domain walls are
wider than $a_{0}$ and then the mis-orientation angle is split across the domain
wall. Also, this method does not count the really small angle domain walls, the
ones with $\theta<12^{\circ}$.

The area of the enclosed domains is used in creating the $N(s_{d})$ histogram.
Fig.5 shows the histogram plot of $N(s_{d})$ for various values of the magnetic
field $b$. The total number of vortices $N_{v}=4096$ and $\Delta=0.02$. At small
and large fields, the histogram can be adequately characterized by a single
parameter, e.g. its half-width. The distribution is relatively narrow with few
large domains. For the intermediate fields, the $N(s_{d})$ exhibits a broad
distribution with substantial weight toward the tail region. This suggests that
more than one parameter is required to characterize these distributions,
especially the excess weight in the tail region.

\begin{figure}[bt]
\epsfxsize=3.2in 
\epsfysize=2.3in 
\includegraphics[width=230pt]{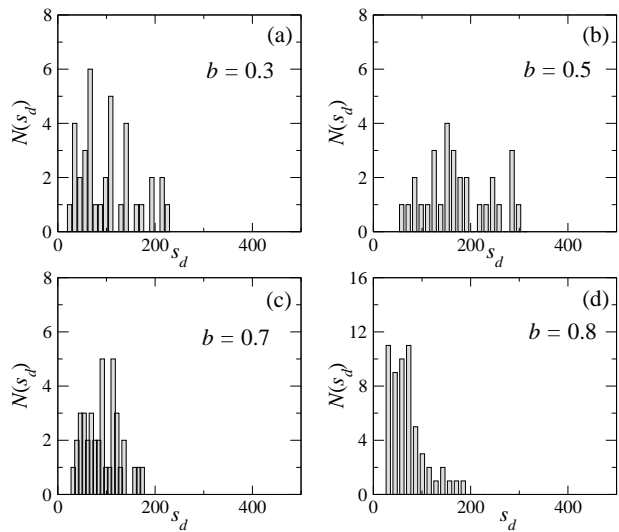} 
\caption{The histogram plot of the domain size distribution $N(s_{d})$, where $s_d$
is in the units of $a_{0}^{2}$. The values of $b$ are (a) 0.3, (b) 0.5, (c) 0.7, 
and (d) 0.8. The $N_{v}=4096$ and $\Delta=0.02$.} 
\end{figure}

\subsubsection{Roughness exponent}

The interaction of the vortex lattice with the quenched impurities leads to
displacement of the vortices from their perfect lattice position. An important
quantity which characterize the change in the position of the vortices is the 
relative displacement correlation, defined as
\begin{equation}
W({\bf r}) = \overline{\left[{\bf u}({\bf r})-{\bf u}(0)\right]^{2}},
\end{equation}
where the overbar represents the average over quenched impurities. The ${\bf
u}({\bf r})$ is the displacement of the vortex relative to its position in the
perfect lattice. The positional order parameter correlations $C_{\bf G}(r)$ can be
expressed in terms of $W(r)$ as $C_{\bf G}(r) \sim e^{-G^{2}W(r)/2}$, where ${\bf
G}$ is one of the reciprocal lattice vectors\cite{gia1}. For the crystalline
state, $W(r)=0$ and $C_{\bf G}(r)=1$. The effect of the quenched impurities is to
increase $W(r)$ and hence reduce the positional order parameter correlations of
the lattice. The structure factor at ${\bf G}$, measured in the neutron scattering
experiments, is related to the Fourier transform of $C_{\bf G}(r)$.

The roughness of an elastic medium is parameterized by the exponent $\zeta$, which
is defined as $W(r)\sim r^{2\zeta}$. In the flat phase of the medium $\zeta<0$, and
in the rough phase $\zeta>0$ (the $\zeta=0$ gives logarithmic roughening with
$W(r)\sim \ln r$). For a $2D$ vortex system, there are three length scales which
emerge in various theories depending upon the displacement $u(r)$:

(1) $r < R_{c}$: In the collective pinning theory\cite{larkin} $R_{c}$ represents
the size of the region in which the vortex lattice is coherently pinned by the
impurities. More precisely, $R_{c}$ is the length scale at which the displacement 
$u(r=R_{c})\sim\xi$. $R_{c}$ is obtained by minimizing the total energy (elastic
energy + pinning energy) and is given by
\begin{equation}
R_{c}\approx\frac{C_{66}\xi}{fn_{p}^{1/2}}.
\end{equation}
The $C_{66}$ is the shear modulus of the vortex lattice and the average pinning
force $f\sim\Delta/r_{p}$, where $r_{p}$ is the range of the pinning potential as
defined in Section II. For the $K_{0}(r/\lambda)$ potential, the field dependence
of the shear modulus have been derived\cite{brandt} and is given as
$C_{66}\approx\frac{\phi_{0}B}{(4\pi\lambda)^{2}}(1-b)^{2}$. In dimensionless
units, the $R_{c}$ becomes 
\begin{equation}
\frac{R_{c}}{a_{0}}\approx\frac{1}{(2\pi)^{3/2}}\frac{\kappa^{2}}{fn_{p}^{1/2}} 
b^{3/2}(1-b)^{2}.
\end{equation}
The $R_{c}$ increases with $b$ and attains a maximum before decreasing as
$b\rightarrow 1$. 

The $r<R_{c}$ regime is often referred as the random force (RF) regime. The
roughness exponent in this regime is given by $\zeta=\frac{4-D}{2}$ for a
$D$-dimensional system. Thus, $\zeta_{RF}^{2D}=1$ and $\zeta_{RF}^{3D}=0.5$ for $2D$
and $3D$ systems, respectively.

(2) $R_{c}<r<R_{a}$: Beyond $R_{c}$, the displacement $u(r)$ continues to grow but
with smaller exponent. $R_{a}$ defines the length scale at which the positional
correlation begins to decay, {\it i.e.}, the displacement $u(r=R_{a})\sim a_{0}$.
Between $R_c$ and $R_a$, the system is in the random manifold (RM) regime. In this
regime, the roughness exponent have been obtained using a Flory type
argument\cite{kardar,nattermann2} which gives $\zeta_{RM}^{2D}=\frac{1}{3}$. A more
refined scaling argument\cite{feigelman} gives $\zeta_{RM}^{2D}=0.4$. For weak
pinning, the length scales $R_{a}$ and $R_{c}$ are related by $R_{a}\sim R_{c}
(a_{0}/\xi)^{1/\zeta_{RM}}$.

(3) $R_{a}<r<\xi_{D}$: Beyond $R_{a}$, $W(r)$ grows as\cite{gia1,nattermann1}
$W(r)\sim\ln^{2}(r)$ as derived through a variational approach and confirmed by
replica symmetric RG\cite{carpentier2}, assuming the lack of dislocations at
these scales\cite{lnlother}. This growth form holds up to the length scale
$\xi_{D}$, at which unbound dislocations appear. For weak pinning, $\xi_{D}\gg
R_{a}$\cite{doussal}:
\begin{equation}
\xi_{D}\sim R_{a} 
\exp\left[c\sqrt{\left(\frac{1}{8}-\sigma_{0}\right)\ln\left(\frac{R_{a}}{a_{0}}\right)}\right]
\end{equation}
where $c$ is a temperature dependent numerical constant and $\sigma_{0}$ is the
impurity strength. For $R_{a}\gg a_{0}$ and low temperatures, $\xi_{D}$ can
become exceedingly large and the system appears similar to the BG phase in $3D$.

(4) $\xi_{D}<r$: Beyond $\xi_{D}$, unbound dislocations lead to exponential decay
of the positional correlation and the system is disordered.

\begin{figure}[bt]
\epsfxsize=3.2in 
\epsfysize=2.3in 
\includegraphics[width=230pt]{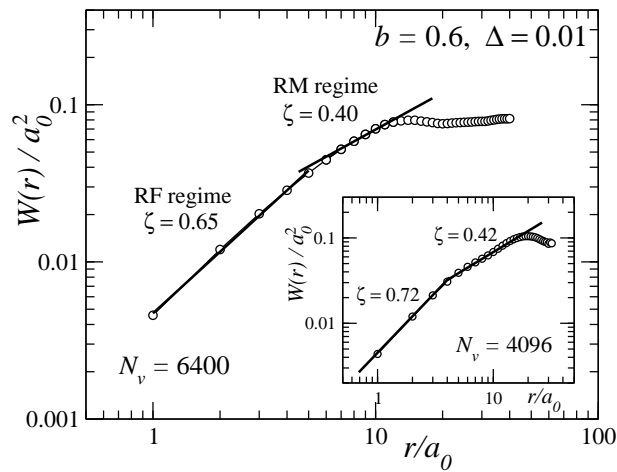} 
\caption{The $W(r)/a_{0}^{2}$ as a function of $r/a_{0}$ for $\Delta=0.01$ for
$N_{v}=6400$. The inset shows the same for $N_{v}=4096$. The $W(r)\sim
r^{2\zeta}$ where the exponent $\zeta$ is shown for the random force (RF) regime
and the random manifold (RM) regime.}
\end{figure}

We have obtained the length scale and the roughness exponent of the vortex
lattice in the DR. The relative displacement correlation $W(r)$ was calculated
using the following procedure. First, a crystalline state with the lattice constant
corresponding to the value of $b$ is constructed using one of the vortex
coordinates ${\bf r}_{0}$ in the real lattice as the origin. The mean square
displacement between the perfect lattice and the underlying real lattice is then
minimized by varying the orientation of the perfect lattice relative to the real
lattice. The $u(r)$ is then computed for each of the vortices. This procedure is
repeated for different ${\bf r}_{0}$'s, and the $W(r)$ is computed by averaging
over all ${\bf r}_{0}$'s.

Fig.6 shows the plot of $W(r)/a_{0}^{2}$ for $b=0.6$ and $\Delta=0.01$ for
$N_{v}=6400$. $W(r)$ for $N_{v}=4096$ is shown in the inset. For these
parameters, even the largest $N_{v}=6400$ system is free of dislocations (see
Fig.2(a)). Due to periodic boundary condition, the length scale probed in the
simulation is half the system size, which for $N_{v}=6400$ and 4096 are 40$a_0$
and 32$a_0$, respectively. $W(r)$ shows an initial power-law increase with an
exponent $\zeta\approx 0.65-0.72$. The system exhibits a crossover around
$r^{*}\approx 4-5a_{0}$ into a regime where the increase of $W(r)$ is slower.
Between $5a_{0}\lesssim r \lesssim 15a_{0}$ the exponent is $\zeta \approx
0.40-0.42$. For $r\gtrsim 16a_{0}$, the growth of $W(r)$ slows down considerably.

It is plausible that $r^{*}=R_{c}$, where the system crosses over from the RF
regime to the RM regime. This can be verified by calculating $R_{c}$ using Eq.(4),
which gives a value $\approx 4a_{0}$ for $b=0.6$, in reasonable agreement with the
value of $r^{*}$. In our system $\xi=0.1\lambda_0$, and at $b=0.6$ the lattice
constant $a_{0}=0.347\lambda_0$. From Fig.6, $u(R_{c})\approx 0.2a_{0}\approx 0.07
\lambda_0$, thus confirming that $u(r=R_{c})\approx\xi$. 

Beyond $R_c$, the length scale $R_a$ of the RM regime is defined as $u(r\!=\!
R_{a})\sim a_{0}$. This translates to $W(r)\approx 0.12a_{0}^{2}$ for $b=0.6$. From
Fig.6, we find that $W(r)$ flattens at $\approx 0.1$ at $r\sim 13a_{0}$ for
$N_{v}=6400$, which suggests that $R_{a}\approx 13a_{0}$ (for $N_{v}=4096$,
$R_{a}\approx 18a_{0}$). Beyond $R_{a}$, the growth of $W(r)$ slows down
considerably which indicates the appearance of the asymptotic regime. Within the
qBG theory\cite{gia1}, $W(r)$ is expected to grow as $\ln^{2}(r)$ in the asymptotic
regime. This behavior unfortunately could not be verified due to insufficient range
of data points. In sum, we identify the $r\sim 1a_{0}-5a_0$ as the RF regime, the
$r\sim 5a_{0}-15a_0$ as the RM regime, and $r>15a_{0}$ as the asymptotic regime.

The value of $\zeta\approx 0.65-0.72$ obtained from the simulation in the RF regime
is smaller than the theoretical prediction for $\zeta_{RF}^{2D}=1$. We speculate
that the interaction $K_{0}(r/\lambda)$ between the vortices in $2D$ increases the
stiffening of the vortex lattice at short distances which leads to weaker
roughening. In the RM regime, the exponent $\zeta\approx 0.40-0.42$ is in good
agreement with the value of 0.4 expected from the scaling argument\cite{feigelman}.
Using $\zeta_{RM}^{2D}=0.4$, the value of $R_{a}\approx 15a_{0}$ is much smaller
than the value $R_{a}\sim R_{c}(a_{0}/\xi)^{1/\zeta_{RM}^{2D}}$. This is possibly
related to the large magnetic field $b=0.6$ used in obtaining $W(r)$. For this
field, $a_{0}$ is comparable to $\xi$, and $R_{c}$ is large compared to smaller
magnetic fields. The RM regime is expected to disappear for $a_{0}=\xi$, and have
been shown in the case of $3D$ system\cite{bogner}.

An interesting outcome of the above analysis of $W(r)$ is that the average domain
size $R_{d}\gg R_{c}$, and hence, the collective pinning theory cannot account for
the appearance of domains in the intermediate fields. The asymptotic regime in
$W(r)$ suggests that the qBG theory is qualitatively correct. Within the qBG
theory, the distribution of dislocations beyond the length scale $\xi_{D}$ is
expected to be homogeneous, unlike the grain boundary formation observed in our
simulation. One possible way to account for the grain boundary formation is to
consider the long range interaction between the dislocations. Since the interaction
between the dislocations is anisotropic, for some values of dislocation density,
the grain boundaries may lead to a lower energy state. This is also supported from
a recent work on dipole systems\cite{tlusty}. At low densities, the dipoles exhibit
a gaseous phase, and their distribution is roughly homogeneous. At higher densities
the phase is characterized by dipoles forming chains or strings. Since the
dislocations of the vortex lattice are in fact dipoles of disclinations, these
results are quite analogous to our identification of a domain regime in the vortex
matter.

\begin{figure}[hbt]
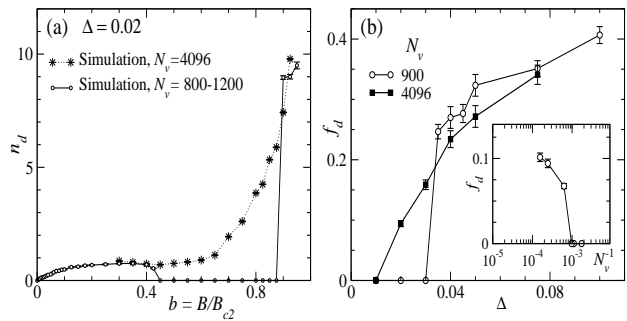
 
\parbox[t]{1.6in}{\includegraphics[height=115pt,width=115pt]{fig7a}}
\parbox[t]{1.6in}{\includegraphics[height=115pt,width=115pt]{fig7b}}
\caption{(a) The topological defect density $n_{d}(b)$ for $\Delta=0.02$ and
$N_{v}=4096$. Also shown is the $n_{d}(b)$ for the smaller system size. (b) The
$f_{d}(\Delta)$ for $b=0.6$.  Inset: The plot of $f_{d}$ as a function of
$N_{v}^{-1}$ for $\Delta=0.02$ and $b=0.6$. There is a critical system size below
which dislocations are not present in the system.}
\end{figure} 

\subsubsection{Defect density}

The three field regimes discussed in the context of the real space configuration
can also be inferred from the behavior of the defect density $n_{d}(b)$. Fig.7(a)  
shows $n_{d}(b)$ for $\Delta=0.02$ and $N_{v}=4096$. The behavior for smaller
system sizes ($N_v$=800-1200) is also shown on the same plot. The $n_{d}(b)$
increases linearly in the low field amorphous regime. Above a crossover field
$b_{l}\approx 0.1$, $n_{d}(b)$ flattens and becomes weakly field dependent in the
DR. For $b\gtrsim 0.6$, $n_{d}(b)$ increases rapidly, and above $b_{h}\approx 0.8$
the system crosses over to the high field amorphous regime. It is possible to
define a length scale $L_{d}\sim n_{d}^{-1/2}$, as the nominal average defect
separation. For $b=0.6$ (domain regime) $L_{d}\approx 3a_{0}$, which is much
smaller than even $R_{c}$ and does not correspond to any feature in the real space
configuration, and reflects the highly inhomogeneous nature of the defect
distribution in the domain regime. On the other hand, in the high-field amorphous
regime $L_{d}\sim a_{0}$, which is also the distance between the defects, thus
reflecting homogeneity of the distribution of defects.

The $n_{d}(b)$ in Fig.7(a) shows strong similarity with the experimental
observation in $2D$ system of magnetic bubbles\cite{shesh}. In Ref.\cite{shesh},
the intermediate regime was interpreted as the hexatic phase and the high field
amorphous regime as the isotropic liquid phase\cite{nelson}. Later
simulation\cite{cha} also suggested a $T=0$ dislocation unbinding transition
driven by disorder. As discussed above, the presence of domain walls (grain
boundaries) in our system suppresses the long range orientational order. This
rules out the possibility of a transition between the hexatic phase and the
isotropic liquid phase as the underlying reason for the rapid increase in
$n_{d}(b)$. However, a rapid crossover, similar to that predicted between the qBG
at low temperatures and vortex liquid at high temperatures\cite{carpentier1}, is
still possible between the DR and the high field amorphous regime, especially at
weaker pinning where the domain size $R_{d}$ is large\cite{comment3}.

For smaller system size ($N_{v}=800-1200$), a topologically ordered phase appears
in the intermediate field range in which $n_{d}=0$ (see Fig.7(a)). This is a finite
size effect, which reflects the sensitivity of the DR to the system size $L$. For
$L<\xi_{D}$, the DR can appear as a topologically ordered state free of
dislocations. This implies that for a given system size, there exists a critical
$\Delta_{c}$ below which dislocations are not favored. This is observed in the
simulation, as shown in Fig.7(b) where the defect fraction $f_{d}(\Delta)$ (number
of defects per vortex) is plotted for $b=0.6$. For the smaller system $N_{v}=900$,
$f_{d}$ goes to zero at $\Delta_{c}=0.03$, and the system exhibits an ordered phase
for $\Delta<\Delta_{c}$. Increasing $N_{v}$ to 4096 reduces $\Delta_{c}$ to 0.01,
and in the asymptotic limit $N_{v}\rightarrow\infty$ (hence, $L\rightarrow\infty$),
we expect $\Delta_{c}\rightarrow 0$. In the DR, $f_d$ does not increase
continuously with increasing $L(\propto N_{v}^{1/2})$ but shows a sharp jump from
the dislocation-free state to the domain state at a characteristic system size as
shown in the inset of Fig.7(b).

\begin{figure}[bt]
\epsfxsize=3.2in 
\epsfysize=2.3in 
\includegraphics[width=230pt]{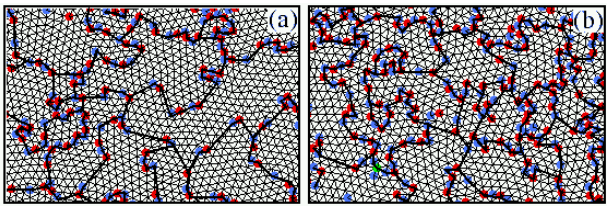} 
\caption{(Color Online) The domains in a region of the simulation box for (a) 
$b=0.7$ and (b) 0.8. The $N_{v}=4096$ and $\Delta=0.02$.}
\end{figure}

\subsubsection{Crossover from DR to high-field amorphous regime}

As discussed above, the $n_{d}(b)$ shows a rapid crossover from the DR to the
high field amorphous regime. To understand the mechanism for this sharp
crossover, we identified the domains and the domain walls between $b=0.5$ and
$b=0.8$ for $\Delta=0.02$. For the intermediate fields $b\approx$ 0.5-0.6, the
grain boundaries are generally smooth and $n_{d}(b)$ is weakly field dependent.
For $b\gtrsim 0.6$, the rapid increase in $n_{d}(b)$ occurs {\em within} the
domain walls. Consequently, the domain wall length increases, which is
accommodated through enhanced roughening of the domain walls. This is evident
from Fig.8(b). The increase in the roughening also facilitates the unbinding of
the dislocations into free disclinations and subsequently drives the crossover
into the VG state. In such a scenario, we conjecture that domain walls undergo
disorder driven roughening transition at the crossover between the DR and the
high field VG. It would be of interest to obtain the domain wall roughening
exponent across the crossover regime.

\begin{figure}[tbh]
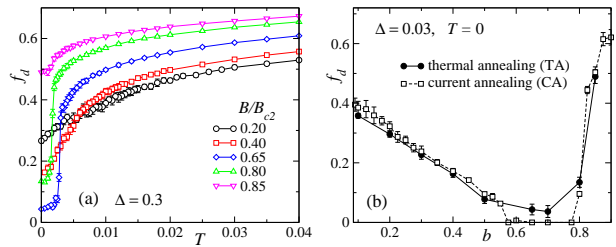

\includegraphics[width=115pt]{fig9a} 
\includegraphics[width=115pt]{fig9b} 
\caption{(Color Online) (a) The defect fraction $f_{d}(T)$ for various magnetic 
fields $b$ with $\Delta=0.03$. The temperature $T$ is decreased slowly from the 
high-$T$ liquid phase. (b) The $f_{d}(b)$ at $T=0$ obtained from thermal annealing 
and current annealing.}
\end{figure} 

\subsection{Finite temperature simulation}

In this section, we compare the current annealing method with the conventional
simulated annealing method, as it is well known that different sample preparation
techniques can result in the vortex system not reaching its equilibrium
configuration\cite{chandran2,jensen}.  In the latter method, the temperature $T$
is reduced from the high temperature liquid phase slowly so as reach thermal
equilibrium at each value of $T$. This method is commonly used to search for the
ground state of disordered systems.

For the thermal annealing, the system was equilibrated for $5\times
10^{4}-1\times 10^{5}$ time steps before averaging over a similar time scale to
calculate the defect fraction $f_{d}(T)$. The number of vortices $N_{v}=900$ and
$\Delta=0.03$. For this system size, the CA method gives a topologically ordered
phase for $b$ between 0.6 and 0.75. Fig.9(a) shows $f_{d}(T)$ for various values
of the magnetic field. As the temperature is lowered, for $b=0.2$ $f_{d}(T)$
decreases monotonically to a finite value with $\frac{df_{d}}{dT}$ slowly
varying. There is no evidence of a transition as a function of the temperature.
With increasing $b$, the slow freezing is replaced by a sharp decrease in
$f_{d}(T)$ at a particular temperature $T_m$, similar to the equilibrium
melting transition. For $b=0.65$, $f_{d}(T)$ at $T_m$ decreases by $\approx
76\%$ of the value above $T_m$. For $b>0.8$, the melting-like transition is
again replaced by slow freezing of the high temperature liquid phase.

In Fig.9(b), $f_{d}(b)$ at $T=0$ obtained by TA is compared with that obtained by
CA. At intermediate fields, the thermally annealed samples exhibit the presence of
dislocations already at these smaller systems sizes. As described above, the
current annealing method requires larger systems sizes to correctly display this
same phenomenon. Otherwise, the two curves track each other very closely over most
of the field range, including the low field slow decay of $f_{d}(b)$ and the rapid
rise at high fields. For the intermediate fields, the $T=0$ configuration obtained
from TA also shows grain boundary formation, similar to that observed from the CA
method.

\section{Conclusion}

We have presented a detailed numerical analysis of the real space configuration of
$2D$ vortex system in the presence of quenched impurities. For weak pinning, the
disordered state in the intermediate field range is inhomogeneous. The majority of
the dislocations in this state are confined to grain boundaries, which form domain
walls between regions of topologically ordered vortex lattice. There are no free
disclinations in the system. This state is referred as the domain state and the
intermediate field range as the domain regime.

The domain size distribution $N(s_{d})$ was calculated in the domain regime.
$N(s_{d})$ shows a broad distribution with a large weight in the tail region at
intermediate fields. Therefore, more than one length scale is required to
properly characterize the domain size distribution in the domain regime. With
increasing $b$, the distribution becomes narrow and the peak shifts toward the
origin. For weak pinning, the size of the domains can become exceedingly large.

The domain regime is bounded by an amorphous regime at low fields and high
fields. The defects in the amorphous regime are separated by the smallest length
scale $\sim a_{0}$ and show homogeneous distribution unlike the grain boundary
formation in the domain regime. The domain regime shows rapid crossover into the
high field amorphous regime. From the changes in the configuration, we identified
the roughening of the domain walls as the plausible mechanism driving the rapid
crossover.

The relative displacement correlation $W(r)$ in the domain state was also
calculated for weak pinning. Three distinct regimes were observed: a random force
regime, a random manifold regime and the asymptotic regime. Crossover from random
force regime to the random manifold regime is found to occur at $R_{c}\sim
4-5a_{0}$. The value of $R_{c}$ agrees with that obtained from the collective
pinning theory. The roughness exponent $\zeta$ in the random manifold regime is
found to $\approx 0.40$, within the range of various theoretical predictions.  

The observation of random manifold and asymptotic regimes {\em within} the domains
for weak pinning suggests that the vortex lattice is correctly described by the qBG
idea, though the exact form of the $W(r)$ could not be ascertained. At length
scales greater than the domain size, the appearance of the domain wall formed by
dislocations is not captured by the quasi-Bragg glass theory.  Therefore, it
remains to be seen whether besides the domain regime the 2D vortex matter supports
a quasi-Bragg glass where the dislocations are homogeneously distributed.

\section{Acknowledgments}

M.C. acknowledges useful discussions with A. K. Grover during the course of the
work. G.T.Z. thanks Y. Fasano, T. Giamarchi, J. Kierfeld, P. Le Doussal, and T.
Nattermann for insightful discussions. M.C. would also like to thank the
University of New Mexico for access to their Albuquerque High Performance
Computing Center. The simulation was performed on the UNM-Alliance Supercluster.
This work was supported by NSF-DMR 9985978.


\begin{references}
\bibitem{comment0} An ensemble of vortex line is referred as the $3D$ vortex system, 
whereas the $2D$ vortex system refers to pancake vortices confined within a plane.
\bibitem{comment1} The term `impurity' is used for the quenched disorder, which
provides the pinning potential for the vortices. The term `disorder' is used to 
identify the state of the vortex configuration in real space.
\bibitem{comment-color} Few vortices have 8 or 4 neighbors in (a) and (f) and is
shown by green dots. We have not attempted to distinguish it in the grayscale for 
the sake of clarity.
\bibitem{comment3} Smaller system size ($N_{v}=800-1200$), for which an ordered
phase appears at intermediate fields for weak pinning, shows a sharp first-order
like transition to the high-field amorphous phase.
\end{references}
\end{document}